# Nonlinear System Identification using Neural Networks and Trajectory-Based Optimization

Hamid Khodabandehlou
EBME Department
University of Nevada, Reno
Reno, Nevada, 89503
hkhodabandehlou@nevada.unr.edu

Mohammad Sami Fadali
EBME Department
University of Nevada, Reno
Reno, Nevada, 89503
fadali@unr.edu

*Abstract*—in this paper, we study the identification of two challenging benchmark problems using neural networks. Two different global optimization approaches are used to train a recurrent neural network to identify two challenging nonlinear models, the cascaded tanks and the Bouc-Wen system. The first approach, quotient gradient system (QGS), uses the trajectories of the nonlinear dynamical system to find the local minima of the optimization problem. The second approach, dynamical trajectory based methodology, uses two different nonlinear dynamical systems to find the connected components of the feasible region and then searches the regions for local minima of the optimization problem. Simulation results show that both approaches effectively identify the model of the cascade tanks and the Bouc-Wen model.

## I. INTRODUCTION

Although engineering applications often require an accurate explicit mathematical model of the system, in many cases such a model is not available. While system models can, in theory, be derived using physical and mathematical principles, deriving such models is difficult in practice [1]. System identification is an alternative approach to derive the model of under test system. System identification uses the input and output measurements of the system to derive its model. The model can be white, grey or black box [1], [2].

Neural networks are one of the most powerful nonlinear system identification tools available. Narendra and Parthasarathy showed that neural networks can effectively identify and control nonlinear dynamical systems [3]. Experimental results show that the neural network can effectively identify the forward and inverse transfer function [4].

Efe and Kaynak extensively studied the identification of nonlinear systems using different neural networks. They studied the application of feedforwad neural networks, radial basis function networks, Runge-Kutta neural networks and adaptive neuro fuzzy inference systems to nonlinear system identification with application to the identification of a robotic manipulator [5]. Other neural networks that have been successful used for nonlinear system identification including Volterra polynomial basis function networks [6], Wavelet networks and Echo state networks [7], and partially recurrent neural networks[8], [9].

Coban proposed the Context Layered Recurrent Neural Network (CLRNN) for identification of linear and nonlinear dynamic systems [9]. CLRNN is a multilayer recurrent neural network with a context layer. The context layer is a feedback from first hidden layer to itself that improves the capability of the network to capture the linear behavior of the system.

Fully Recurrent Neural Networks (FRNN) are recurrent neural networks in which all the nodes of the hidden layer are connected. FRNN's can effectively identify linear and nonlinear models. However, their complicated structure makes their training difficult and slow [11],[12],[13]. Backpropagation Through Time (BPTT) is an alternative to traditional error backpropagation for training recurrent neural networks. BPTT represents recurrent neural network as a multilayer feedforward network, then tunes its weights using backpropagation [14],[15][16]. BPTT is computationally expensive and its computational time increases drastically with the size of the training dataset. It also and suffers from vanishing gradient problem [17]. Regularization of the neural network weights is an approach to cope with the vanishing gradient problem [17]. Sutskever proposed Hessian free optimization to train recurrent neural networks and overcome the vanishing gradient issue [18].

Williams and Zipser proposed Real Time Recurrent Learning (RTRL) for training recurrent neural networks. The RTRL does not need a precisely defined training interval but suffers from huge computational complexity in large applications [19]. Generalized Long Short Term Memory (LSTM) is another approach for training second order recurrent neural networks [20]. The method is applicable to a wide range of second order recurrent networks and has better performance than traditional LSTM.

Lu et. al. used low rank factorization to inspect redundancies in recurrent neural networks [21]. They argued that using structured matrices and shared low-rank factors can effectively reduce the number of parameters of the standard LSTM without significantly increasing error.

Although a wide variety of algorithms have been used to train feedforward and recurrent neural networks, there are

This paper is based upon work supported by the National Science Foundation under Grant No. IIA-1301726.

promising optimization approaches that have not been used for training and that have the potential to provide better system identification results. This paper explores the use of two trajectory-based methodologies for nonlinear system identification: the quotient gradient method and the dynamical trajectory-based approach.

The quotient gradient method is a trajectory-based methodology to find the possible feasible solutions of the constraint satisfaction problem. Quotient gradient uses the trajectories of stable nonlinear dynamical system to find the solutions of the original constraint satisfaction problem [22].

The Dynamical Trajectory Based approach (DTB) is another global optimization approach that is applicable to general constrained optimization problems. DTB uses the trajectories of two nonlinear dynamical systems, i.e. Projected Gradient System (PGS) and quotient Gradient System (QGS) to find disjoint components of the feasible region of the optimization problem and search those disjoint components for possible solutions of the optimization problem [23].

In this study, we use quotient the gradient method and DTB to train recurrent neural network and evaluate the performance of the networks on two of the challenging nonlinear system identification benchmarks. The first is the cascaded tank model, which is difficult to identify due to saturation and overflow in the tanks. The second is the Bouc-Wen model. The Bouc-Wen model is highly nonlinear model with hysteretic behavior, which makes it challenging benchmark for system identification

The remainder of this paper is organized as follows: Section II presents the neural network structure. Section III describes the quotient gradient method and section IV describes the dynamical trajectory based optimization approach. Section V describes the benchmark systems and Section VI presents simulation results.

## II. NEURAL NETWORK

In this study, we use a fully recurrent neural network with one hidden layer. The structure of the neural network is shown in Fig. 1.

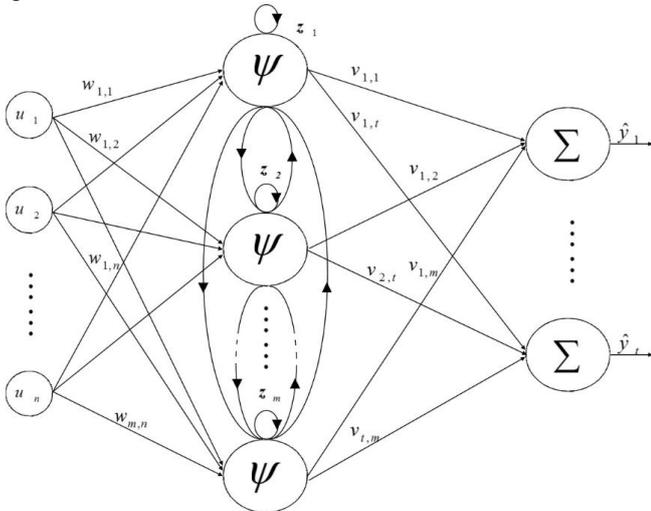

Fig. 1. Internal structure of the neural network

For a network with $n$ inputs, $m$ hidden layer nodes, and $t$ outputs, the input, internal state and output vector of the network, respectively, are:

$$\boldsymbol{u}(k) = [u_1(k) \quad \ldots \quad u_n(k)]^T$$
$$\boldsymbol{z}(k) = [z_1(k) \quad \ldots \quad z_m(k)]^T \quad (1)$$
$$\hat{\boldsymbol{y}}(k) = [\hat{y}_1(k) \quad \ldots \quad \hat{y}_t(k)]^T$$

The governing equation of the network is

$$\boldsymbol{z}(k) = \boldsymbol{\psi}(W\boldsymbol{u}(k) + S\boldsymbol{z}(k-1))$$
$$\hat{\boldsymbol{y}}(k) = V\boldsymbol{z}(k) \quad (2)$$

where $W_{m \times n}$, $S_{m \times m}$ and $V_{t \times m}$ are weight matrices. The network is trained to determine the optimal value of the weight matrices. The cost function for training is the sum of squared errors (SSE)

$$SSE = \sum_{k=1}^{N} \boldsymbol{e}(k)^T \boldsymbol{e}(k)$$
$$= \sum_{k=1}^{N} (\hat{\boldsymbol{y}}(k) - \boldsymbol{y}(k))^T (\hat{\boldsymbol{y}}(k) - \boldsymbol{y}(k)) \quad (3)$$

$N$ is the number of training samples. $\boldsymbol{y}$ is the measured output and $\hat{\boldsymbol{y}}$ is the output of neural network.

## III. QUOTIENT GRADIENT METHOD

The problem of training neural networks is a nonlinear optimization problem. QGS is a nonlinear dynamical system to find local possible feasible solutions of the constraint satisfaction problem. Quotient gradient method transforms the constraint satisfaction problem into an unconstrained minimization problem, then uses QGS to find its local minima, which are the feasible solutions of the constraint satisfaction problem. Consider the following CSP:

$$C_I(y) < 0$$
$$C_E(y) = 0, y \in R^{n-l} \quad (4)$$

where $C_I$ are inequality constraints and $C_E$ are equality constraints. Equality and inequality constraints are assumed smooth to guarantee the existence of the solution. This CSP can be transformed to the unconstrained minimization problem

$$\min_x f(x) = \frac{1}{2} \|H(x)\|^2, \quad x = (y, s) \in R^n \quad (5)$$

$$H(x) = \begin{bmatrix} C_I(y) + \hat{S}^2 \\ C_E(y) \end{bmatrix} \in R^m, \hat{S}^2 = (s_1^2, \ldots, s_l^2)^T \quad (6)$$

where $\hat{S}$ is set of slack variables to transform inequality constraints to equality constraints. Lee and Chiang argued that local minima of the unconstrained optimization problem are possible feasible solutions of the original CSP. They introduced the QGS and showed that its equilibrium points are local minima of the unconstrained minimization problem, which are possible feasible solutions of the CSP. QGS for the unconstrained optimization problem is defined as

$$\dot{x} = F(x) = -\nabla f(x) \coloneqq -D_x H(x)^T H(x) \quad (7)$$

QGS is a completely stable system, therefore, integrating QGS from any arbitrary point leads to an equilibrium point, which is local minimum of (5), and a feasible solution of (4).

After finding the first equilibrium point, QGS must escape from its basin of attraction and enter the basin of attraction of another equilibrium point. This can be done by backward integration of QGS in time until it reaches an unstable point. Therefore, the process of finding local minima of (5) becomes a series of forward and backward integrations of QGS.

To train a neural network, we optimize the cost function (SSE) to find the optimal values of the weight matrices. To optimize SSE using the quotient gradient method, the weight matrices are partitioned as

$$V = \begin{bmatrix} v_1^T \\ \vdots \\ v_m^T \end{bmatrix}_{t \times m} \quad W = \begin{bmatrix} w_1^T \\ \vdots \\ w_m^T \end{bmatrix}_{m \times n} \quad S = \begin{bmatrix} s_1^T \\ \vdots \\ s_m^T \end{bmatrix}_{m \times m} \quad (8)$$

The vector of network parameters $x$ is defined as
$$x = [x_i]_{n_p \times 1} = [v_1, .., v_m, w_1, ..., w_m, s_1, ..., s_m]^T$$
$$n_p = m^2 + m \times (n + t) \quad (9)$$

Using the vector $x$, the training set can be rewritten as
$$h(x) = [h_i(x)], i = 1, 2, ..., N$$
$$h_i(x) = V\psi(Wu(i) + Sz(i-1)) - y(i) \quad (10)$$

The QGS for training neural network can be constructed as
$$\dot{x} = -f(x) = -D_x h(x)^T h(x) \quad (11)$$

where
$$D_x h(x) = \begin{bmatrix} \frac{\partial h_1(x)}{\partial x} \\ \vdots \\ \frac{\partial h_N(x)}{\partial x} \end{bmatrix}_{N \times n_p} \quad (12)$$

Therefore, the problem of finding optimal values of weight matrices becomes: (1) forward integration of the QGS until it reaches an equilibrium point (2) backward integration of QGS until it reaches an unstable equilibrium point and (3) forward integration of QGS from unstable point until it reaches another stable equilibrium point. This process continues until all the equilibrium points of the QGS are determined. The equilibrium point with the lowest cost is the global optimum of the optimization problem. The eigenvalues of the Jacobian matrix are a measure of instability of the points during backward integration [22][24].

## IV. DYNAMICAL TRAJECTORY BASED APPROACH

The quotient gradient method provides a systematic approach to find the local minima of the optimization problem. However, in most of the optimization problems, the feasible region, $M$, is union of disjoint connected components

$$M = \bigcup_i M_i \quad (13)$$

Dynamical trajectory based optimization provides a systematic method to find the connected components of the feasible region and search those components for local minima. The approach has two phases: the quotient gradient system (QGS) to find connected components of the feasible region, and the projected gradient system (PGS) to search the feasible components for local minima [25]. Consider the following constrained minimization problem

$$\begin{aligned} \min & f(x) \\ \text{s.t.} & \ h(x) = 0 \end{aligned} \quad (14)$$

Assume $f(x) \in C^2(R^n, R^m)$ and $h(x)$ to be smooth to guarantee the existence of the solution. The inequality constraints can be incorporated in the optimization problem by introducing slack variables. The feasible region of the optimization problem is defined as

$$M := \{x \in R^n : h(x) = 0\} \quad (15)$$

### A. PGS Phase

Searching connected components of the feasible region for local minima is an essential part of the DTB approach. DTB uses PGS to find multiple local minima in connected components of the feasible region. PGS is a stable nonlinear dynamical system whose equilibrium points are local minima of the optimization problem. After finding one local optimal solution, the trajectories of PGS can be used to move away from current local minimum and move toward another local minimum in the current component of the feasible region. The PGS is defined as

$$\dot{x} = F(x) = -\nabla f_{\text{proj}}(x), \quad x \in M \quad (16)$$

where $\nabla f_{\text{proj}}(x)$ is orthogonal projection of $\nabla f(x)$ on the tangent space of the feasible region. It can be shown than when $Dh(x) = \partial h(x)/\partial x$ is nonsingular, $\nabla f_{\text{proj}}(x)$ is defined as

$$\nabla f_{\text{proj}}(x) = \big(I - Dh(x)^T(Dh(x)Dh(x)^T)^{-1}Dh(x)\big)\nabla f(x) \quad (17)$$

Every PGS trajectory converges to one of its stable equilibrium points, which is also a local optimum of (14). After finding one local minimum, PGS needs to escape from stability region of that local minimum and enter the stability region of another local minimum in the current component of the feasible region. This is achieved by backward integration of PGS until reaching a saddle point. Then by forward integration of PGS moves it towards another local optimal solution. By repeating this process, PGS finds all the local optimal solutions in the current component of the feasible region. The next step is moving toward another component of the feasible region, which is done in the QGS phase.

### B. QGS Phase

To explore all the components of the feasible region, DTB approach needs to escape from current component and move to another component of the feasible region. DTB uses the trajectories of a nonlinear dynamical system to do this. The nonlinear dynamical system to explore the components of the feasible region, the QGS, is

$$\dot{x} = -Dh(x)^T Dh(x) \quad (18)$$

where $Dh(x)$ is the Jacobian of $h$ at $x$. Every QGS trajectory converges to one of its stable equilibrium manifolds and every stable QGS equilibrium manifold corresponds to a connected component of the feasible region. Therefore, to approach a connected component of the feasible region, the QGS is integrated until it reaches an equilibrium point. To escape from the current component of the feasible region and move toward another feasible component, the QGS is integrated backward in time until it reaches an unstable point, then integrated forward in time until it reaches another component of the feasible region. By invoking PGS and QGS phases repeatedly, the DTB approach finds multiple components of the feasible region and locates local optimal solutions. The local optimal solution with the lowest cost is the global optimal solution of the optimization problem.

Although training neural networks is an unconstrained optimization problem, constraints are needed for defining QGS. We define the constraints of the optimization problem as upper and lower bounds on the neural network weights. The constraints are written in terms of the network parameters of (9) as

$$|x_i| \leq l_i, \quad i = 1, \dots, n_p \quad (19)$$

By adding slack variables, $\boldsymbol{s}^T = [s_1, \dots, s_{n_p}]$, inequality constraints can be written as equality constraints

$$h_i(\boldsymbol{x}) = x_i^2 - l_i^2 + s_i^2 = 0, \quad i = 1, \dots, n_p \quad (20)$$

The augmented vector of parameters is defined as

$$\boldsymbol{o} = [\boldsymbol{x}\ \boldsymbol{s}]^T{}_{(2 \times n_p) \times 1} \quad (21)$$

Equation (14) can be rewritten in terms of $\boldsymbol{o}$ as

$$\begin{aligned} \min f(\boldsymbol{o}) \\ \text{s.t.} \ \boldsymbol{h}(\boldsymbol{o}) = \boldsymbol{0} \end{aligned} \quad (22)$$

The constraint set $D\boldsymbol{h}(\boldsymbol{o}) = [\partial h_i(\boldsymbol{o})/\partial \boldsymbol{o}]^T_{(n_p) \times (2 \times n_p)}$ is always nonsingular and the PGS and QGS for training neural network are

PGS:
$$\dot{\boldsymbol{o}} = -\big(I - D\boldsymbol{h}(\boldsymbol{o})^T(D\boldsymbol{h}(\boldsymbol{o})D\boldsymbol{h}(\boldsymbol{o})^T)^{-1}D\boldsymbol{h}(\boldsymbol{o})\big)\nabla f(\boldsymbol{o}) \quad (23)$$

QGS:
$$\dot{\boldsymbol{o}} = -D\boldsymbol{h}(\boldsymbol{o})^T \boldsymbol{h}(\boldsymbol{o}) \quad (24)$$

This reduces training neural networks to repeated invoking of PGS and QGS until a stopping criterion is satisfied. To avoid the effect of upper and lower bound on the final solution, the upper and lower bounds can be chosen arbitrarily large [25].

## V. BENCHMARK SYSTEMS

### A. Cascade tank model

The cascade tank system is a challenging nonlinear benchmark for system identification. The system consist two tanks with a pump. Fig. 2. Shows the structure of the cascade tank model. The pump feeds water into the upper tank and the lower tank has a free outlet. The system is not highly nonlinear during normal operation. However, with large water flow into the upper tank, overflow can occur in the upper tank. This overflow acts as an input-dependent process noise. Without overflow, the cascade tank is governed by

$$\begin{aligned} \dot{x}_1(t) &= -k_1\sqrt{x_1(t)} + k_4 u(t) + w_1(t) \\ \dot{x}_2(t) &= k_2\sqrt{x_1(t)} - k_3\sqrt{x_2(t)} + w_2(t)y(t) \\ y(t) &= x_2(t) + e(t) \end{aligned} \quad (25)$$

where $u(t)$ is the pump voltage, $x_1(t)$ and $x_2(t)$ are the states of the cascade tank system, $w_1(t), w_2(t)$ and $e(t)$ are noise and $k_1, k_2, k_3$ and $k_4$ are system constants. $y(t)$ is the system output, i.e., the output of the second tank.

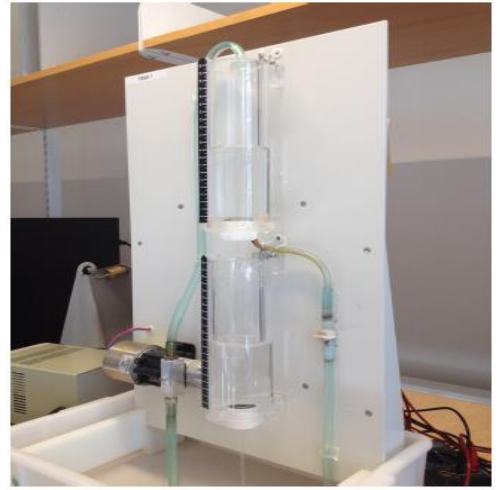

Fig. 2. Cascade tank model structure

The input is a multisine signal with frequencies from 0 to 0.0144 Hertz. Lower frequency inputs have larger amplitude than higher frequency inputs. Sampling time is $T_s = 4s$. A capacitive water level sensor is used to measure the water level and is assumed to be a part of the system [26].

### B. Bouc-Wen model

The Bouc-Wen model is a widely studied hysteresis model in mechanical and civil engineering [28], [29]. A hysteretic system has multiple stable equilibrium points; therefore output can change based on the input history which makes the system more complicated for analysis and design [27]. The Bouc-Wen oscillator is governed by

$$m_L \ddot{y}(t) + r(y, \dot{y}) + z(y, \dot{y}) = s(t) \quad (26)$$

where $s(t)$ is the input, i.e. the external force, $y(t)$ is the displacement and $m_L$ is the mass constant. $r(y, \dot{y})$ is the total restoring force and $z(y, \dot{y})$ is a history dependent nonlinear term that determines the hysteretic property of the system. The static restoring force is

$$r(y, \dot{y}) = k_L y + c_L \dot{y} \quad (27)$$

where $k_L$ linear stiffness coefficient and $c_L$ is viscous damping coefficient. $z(y, \dot{y})$ is

$$\dot{z}(y, \dot{y}) = \alpha|\dot{y}| - \beta(\gamma|\dot{y}||z|^{v-1} + \delta \dot{y}|z|^v) \quad (28)$$

$\alpha, \beta, \gamma, v$ and $\delta$ are Bouc-Wen parameters that determine the shape and smoothness of the hysteresis loop. Table. I. shows the parameter values used in this study.

Table. I. System parameters

| Parameter | Value |
|---|---|
| $m_L$ | 2 |
| $c_L$ | 2 |
| $k_L$ | $5 \times 10^4$ |
| $\alpha$ | $5 \times 10^4$ |
| $\beta$ | $10^3$ |
| $\gamma$ | 0.8 |
| $\delta$ | $-1.1$ |
| $v$ | 1 |

## VI. SIMULATION RESULTS

We use the quotient gradient method and the DTB approach to train a neural network for identification of benchmark problems and compare the results.

### A. Cascade tanks

For a fair comparison between QGS and DTB, the neural networks structure, including training data, input vector, hidden layer activation function and number of hidden layer nodes is the same in all the simulations. All the network parameters were initialized with random values from a zero-mean normal distribution with standard deviation $\sigma^2 = 0.1$. The optimal number of hidden layer nodes was found to be $m = 9$ and the network input is

$$\boldsymbol{u}(k) = [1, s(k), \ldots, s(k-9), y(k-1), \ldots, y(k-9)]^T \quad (29)$$

The target value for network output is $y(k)$. Fig .3 shows the validation data and the output of the trained networks and Fig. 4 shows the validation error of the networks.

Although both networks have a very good performance in the identification of the nonlinear model, the dynamical trajectory based method has a lower mean squared error than the QGS trained network for validation data. The mean squared error of the DTB trained network is $MSE = 0.0264$ while the mean squared error of QGS trained network is $MSE = 0.0312$.

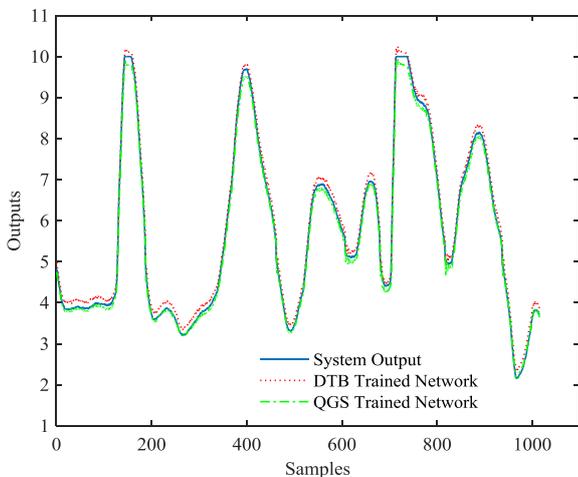
Fig. 3. Validation outputs for the networks trained with DTB and QGS.

Although both approaches successfully identify the model of cascaded tanks, the dynamical trajectory based approach gives slightly better results. This is because it more accurately determines the local minima of the optimization problem.

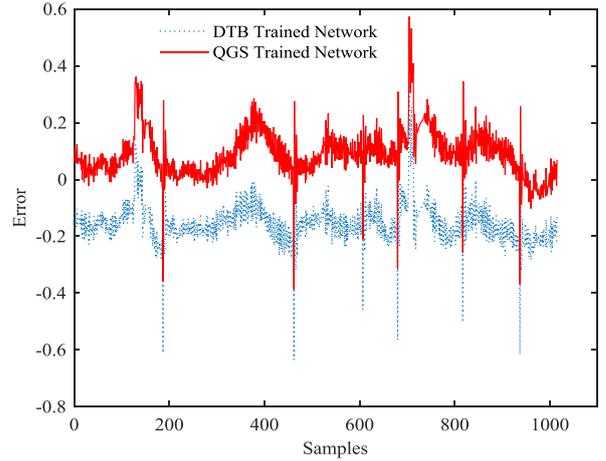
Fig. 4. Validation error for the networks trained with DTB and QGS

### B. Bouc-Wen Model

For a fair comparison, all the parameters for QGS and DTB approach are assumed the same. The network parameters were initialized with random values from zero-mean normal distribution with standard deviation of $\sigma^2 = 0.1$. The optimal number of hidden layer nodes was found to be $m = 7$ and the neural network's input vector is assumed to be

$$\boldsymbol{u}(k) = [s(k), \ldots, s(k-5), y(k-1), \ldots, y(k-5)]^T \quad (30)$$

The target value for network output is $y(k)$. Fig .5 shows the validation output and the output of the trained networks. Fig. 6 shows the validation error of the networks and shows that the dynamical trajectory based trained network has better performance than the QGS trained network. The mean squared error of the DTB trained network is $MSE = 2.3 \times 10^{-8}$. while the mean squared error of the QGS trained network is $MSE = 6.1 \times 10^{-8}$.

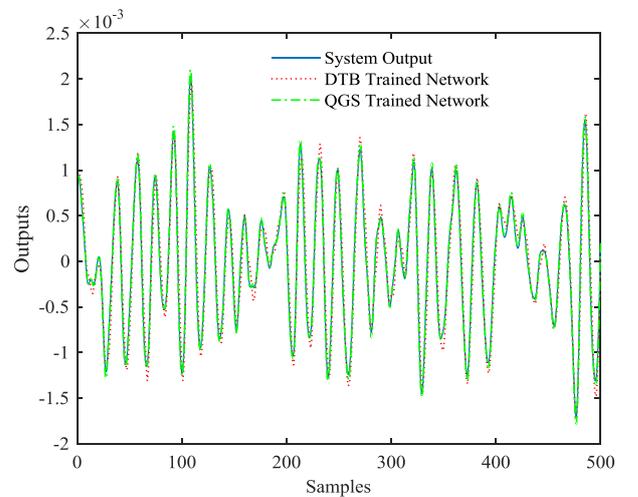
Fig. 5. Test outputs for the networks trained with DTB and QGS.

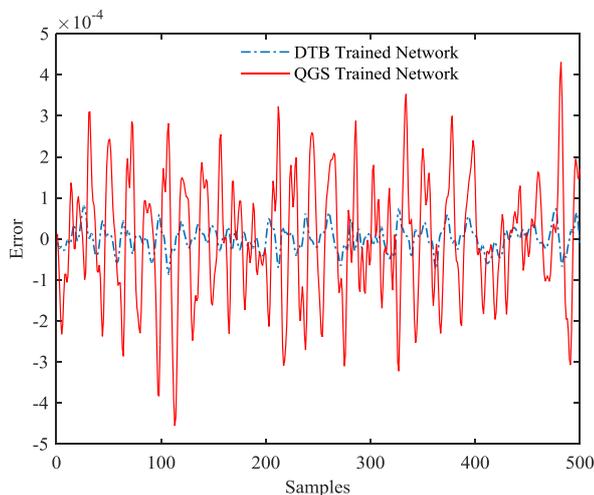

Fig. 6. Test error of the networks trained with DTB and QGS.

## VII. CONCLUSION

In this study, we used two trajectory-based optimization approaches to train artificial neural networks for the identification of two nonlinear system identification benchmark problems: cascaded tanks and the Bouc-Wen model. Both approaches use trajectories of nonlinear dynamical systems to find optimal value of the neural network weights and were able to train neural network and efficiently identify nonlinear system models. Although both approaches successfully identify the nonlinear models, the dynamical trajectory based approach has better performance at the expense of longer training time. Future work will include designing a neural network based controller for nonlinear systems.